\documentclass[12pt]{article}
\usepackage{amssymb}
\usepackage{amsmath}
\usepackage{amstext}
\usepackage{graphicx,epsfig}
\usepackage{epsfig}
\usepackage{verbatim} 
\usepackage{fancybox}
\usepackage{color}
\usepackage{ulem}
\usepackage{enumitem}
\usepackage{subfigure}
\usepackage{bbm}
\usepackage{parskip}
\usepackage{dsfont}
\usepackage[numbers,sort&compress]{natbib}
\usepackage{ytableau}
\usepackage{cancel}
\usepackage{verbatim}

\usepackage{fancyhdr}	
\usepackage{indentfirst}	
\usepackage{amsmath,amssymb,latexsym,epsf,epsfig,graphicx,tikz}
\usepackage[utf8]{inputenc} 
\usepackage{empheq}		
\usepackage{amsthm}
\usepackage{amsmath}
\usepackage{eucal}
\usepackage{mathtools}
\usepackage{mathrsfs}
\usepackage{yfonts}
\usepackage{array}		
\usepackage[utf8]{inputenc}

\usepackage{tikz}
\usetikzlibrary{arrows,shapes}
\usetikzlibrary{trees}
\usetikzlibrary{matrix,arrows} 				
\usetikzlibrary{positioning}				
\usetikzlibrary{calc,through}				
\usetikzlibrary{decorations.pathreplacing}  
\usepackage{pgffor}							

\usetikzlibrary{decorations.pathmorphing}

\newcommand{\upleft}[2]{\prescript{(#1)}{}{\! #2}} 

\newcommand{\Comment}[1]{{}}
\definecolor{darkblue}{rgb}{0.95,0.05,0.05}
\definecolor{reddish}{rgb}{0.95, 0.05, 0.05}
\usepackage[linktocpage=true]{hyperref}
\hypersetup{
colorlinks=true,
citecolor=darkblue,
linkcolor=reddish,
urlcolor=darkblue,
pdfauthor={},
pdftitle={},
pdfsubject={}
}

\newcommand{\nn}{\nonumber \\ }

\def\({\left(}
\def\){\right)}

\newcommand{\beq}{\begin{equation}}
\newcommand{\eeq}{\end{equation}}
\newcommand{\be}{\begin{equation}}
\newcommand{\ee}{\end{equation}}
\newcommand{\bea}{\begin{align}}
\newcommand{\eea}{\end{align}}

\def\gsim{ \lower .75ex \hbox{$\sim$} \llap{\raise .27ex \hbox{$>$}} }
\def\lsim{ \lower .75ex \hbox{$\sim$} \llap{\raise .27ex \hbox{$<$}} }

\def\beq{\begin{eqnarray}}
\def\eeq{\end{eqnarray}}

\def\p{{\cal P}}

\def\L*{{\cal L}_*}
\def\L{\mathcal{L}}
\def\({\left(}
\def\){\right)}

\def\nn{\nonumber}
\def\p{\partial}

\def\stu{St\"uckelberg }
\def\p{\partial}

\def\<{\langle}
\def\>{\rangle}

\def\xyma{\xymatrix@M.7em}
\def\xymas{\xymatrix@M.1em}
\newcommand{\ba}{\begin{eqnarray}}
\newcommand{\ea}{\end{eqnarray}}

\usepackage{bbold}

\usepackage{tikz}
\usetikzlibrary{decorations}
\pgfdeclaredecoration{complete sines}{initial}
{
    \state{initial}[
        width=+0pt,
        next state=upsine,
        persistent precomputation={\pgfmathsetmacro\matchinglength{
            \pgfdecoratedinputsegmentlength / int(\pgfdecoratedinputsegmentlength/\pgfdecorationsegmentlength)}
            \setlength{\pgfdecorationsegmentlength}{\matchinglength pt}
        }] {}
    \state{upsine}[width=\pgfdecorationsegmentlength,next state=downsine]{
        \pgfpathsine{\pgfpoint{0.25\pgfdecorationsegmentlength}{0.5\pgfdecorationsegmentamplitude}}
        \pgfpathcosine{\pgfpoint{0.25\pgfdecorationsegmentlength}{-0.5\pgfdecorationsegmentamplitude}}
    }
    \state{downsine}[width=\pgfdecorationsegmentlength,next state=upsine]{
        \pgfpathsine{\pgfpoint{0.25\pgfdecorationsegmentlength}{-0.5\pgfdecorationsegmentamplitude}}
        \pgfpathcosine{\pgfpoint{0.25\pgfdecorationsegmentlength}{0.5\pgfdecorationsegmentamplitude}}
}
    \state{final}{}
}

\title{}
\author{}

\numberwithin{equation}{section}

\begin{document}
%
\renewcommand{\thefootnote}{\fnsymbol{footnote}}
~
%
%
\begin{center}
{\Large\bf{Boundary Terms for Massive General Relativity }} 
\end{center} 
 \vspace{0.9truecm}
\thispagestyle{empty} 
\centerline{
{\large {Gregory Gabadadze${}^{\rm a}$}}
{\large {and David Pirtskhalava${}^{\rm b}$}}
}

\vspace{.7cm}

\centerline{{\it $^a$Center for Cosmology and Particle Physics, Department of Physics}}
\centerline{{\it New York University, New York, NY, 10003, USA}}
\vskip 0.15cm
\centerline{{${}^{\rm b}$\it Theoretical Physics Group, Blackett Laboratory }}
\centerline{{\it Imperial College, London SW7 2AZ, UK}}

 \vspace{1.cm}
\begin{abstract}
\noindent
 It is well-known that the presence of a spacetime boundary
requires the conventional Einstein-Hilbert (EH) action to be supplemented by the 
Gibbons-Hawking (GH) boundary term in order to retain  the standard  variational 
procedure.~When the EH action is amended by the diffeomorphism-invariant graviton mass and potential terms, 
it naively appears that no further boundary terms are needed since all the new fields of massive gravity 
enter the action with the first derivative.  However, we show that such a formulation would be inconsistent, even when 
the bulk action is ghost-free. The theory is well-defined only after introducing novel boundary counterterms, which dominate 
over the GH term in the massless limit and cancel the problematic boundary terms induced by the bulk action. The number of boundary 
counterterms equals the number of total derivatives one could construct in the bulk 
using positive powers of two derivatives  of the  longitudinal mode of the massive graviton.

\end{abstract}

\newpage

\setcounter{tocdepth}{2}
\newpage
\renewcommand*{\thefootnote}{\arabic{footnote}}
\setcounter{footnote}{0}

\tableofcontents

\section{Introduction and summary}
\noindent
The conventional action\footnote{We use the `mostly plus' metric convention throughout this manuscript.},  -$\int_{\it M} (\partial\phi )^2$,   for a scalar field $\phi$ on a $D$-dimensional 
flat spacetime ${\it M}$ endowed with a 
boundary ${\partial {\it M}}$  does not require a boundary term  for the standard variational  principle to be well-defined.
If, on the other hand, a ``less-conventional''  form of the action were used, $\int_{\it M} \phi \square\phi $, 
one would need to introduce a boundary term, $\int_{\partial {\it M}} \phi (\mathbf{n}\cdot\p) \phi $, where $\mathbf{n}$ denotes the \textit{outward} normal to the boundary.~The latter term is required to avoid specifying the boundary value of the first derivative of the field's variation, induced by the bulk action containing second derivatives of $\phi$.

\vskip 0.15cm
\noindent
In the conventional covariant formulation of Maxwell's theory, no boundary terms are needed as the vector field appears in the action with first derivatives only.~However, the Einstein-Hilbert (EH) action, written in its standard form $\int d^Dx\, \sqrt{-g}\, R$, contains terms with second derivatives acting on the metric, and requires the Gibbons-Hawking (GH) term to be added on the spacetime boundary \cite {Gibbons:1976ue}. 

\vskip 0.15cm
\noindent
The diffeomorphism-invariant theory of massive gravity  \cite {deRham:2010ik,deRham:2010kj}  amends the EH action by the  
mass and nonlinear potential terms of the form \cite{deRham:2010kj} 
\be
S \propto \int_{\it M} d^Dx\, \sqrt{-g}\, U \left (\sqrt { g^{-1} \mathbb{f} } \right ) \,,
\label{massU}
\ee
where $U$ is a special polynomial of the square root of a matrix that is the product of the inverse of the metric 
$g_{AB}~(A,B= 0,...,3,5,...,D$) and the``fiducial'' metric $\mathbb{f}_{AB} = \partial_A \Phi^I \partial_B \Phi^J \eta_{IJ}$\, ($I,J =  0,...,3,5,...,D$ are the \textit{internal}-space indices labelling the space-time scalars $\Phi^I$).~The action for these scalars is a 
special sigma model  with $ISO(1,D-1)$ internal global symmetry, that does not lead to loss of unitarity and contains the scalar fields only through first derivatives $\partial\Phi$. Moreover, there are no derivatives acting on the metric tensor 
in (\ref {massU}); hence, one would naively  expect that no further boundary terms are needed  when 
the EH action is amended by (\ref {massU}).

\vskip 0.15cm
\noindent
Nevertheless, we will show that in a general diff-invariant formulation of massive gravity on a manifold with a boundary, certain  
additional boundary terms are required.  This claim seems   
counterintuitive due to the fact that the new fields enter only through first derivatives;  however, there is a more detailed consideration which makes things clear.~The scalar fields $\Phi^I$ encode  the $D-1$ physical degrees of freedom of a massive graviton 
in a gauge in which $g_{AB}$ encodes the remaining $D(D-3)/2$ dynamical modes (as in massless gravity).~One of them,  the longitudinal mode, enters  $\Phi^I$ with a derivative, $\Phi^I\supset \delta_A^I \partial^A \Pi$\,; while in the 
full theory  $\Pi$ is a gauge mode and carries no independent meaning, in  a special high-energy 
limit of the theory (which we will refer to as the \textit{decoupling limit}) it becomes a gauge-invariant degree of freedom and becomes physical. 
If so, its action should be well-defined.~In order to see whether this is indeed the case, consider the Minkowski background of the theory, $\langle g_{AB} \rangle =\eta_{AB}$
and $\langle \partial_A\Phi^I \rangle = \delta_A^I$, which is a solution to the equations of motion of massive gravity.~In  the linearized 
approximation, the nonlinear potential (\ref {massU})  reduces to the Fierz-Pauli  (FP) term~\cite{Fierz:1939ix}
\be
(h_{AB}  - \partial_A V_B - \partial_B V_A)^2 - (h^A_A -  2 \partial^A V_A)^2, 
\label{introFP}
\ee
where  $h_{AB} \equiv g_{AB} - \eta_{AB}$  and  $V_B \equiv  \delta_B^ I\, \eta_{IJ}\, ( \Phi^J - X^A\delta_A^J)$
are the fluctuations of the fields involved. The FP term is known to be the only quadratic  term  free  of ghosts, 
tachyons, and other instabilities \cite{VanNieuwenhuizen:1973fi}.   The necessary condition for the absence of ghosts is that the equations of motion for
the longitudinal mode, $V_B = \partial_B \Pi$, contain no more  than two  time derivatives. This is certainly  the case if one ignores the total derivative that emerges from (\ref {introFP}) 
\be
\left ( \partial_A \partial_B \Pi  \right  )^2 - \left ( \partial^A \partial_A \Pi  \right  )^2 \,.
\label{ddPi}
\ee
This term dominates over the others in the massless limit (due to the coefficient that is not shown here, but see discussions in \cite {deRham:2010ik} and in Sec.~\ref{w/obound}). While harmless in an unbounded spacetime and for the fields that 
decay fast enough at infinity, this term would give rise to more than two time  derivatives in the effective action for boundary degrees of freedom, if a background manifold with a boundary were considered. Hence, for consistency, a boundary can only be introduced with a counterterm localized on it, to cancel the 
otherwise problematic pullback of (\ref {ddPi}).

\vskip 0.15cm
\noindent
Similar considerations apply to nonlinear terms in (\ref {massU}): these terms were built by requiring that the action for the longitudinal mode, with all other fields set to zero, is just a total derivative up to order D in non-linearities, and is zero at higher orders \cite {deRham:2010ik}. For each of these  
total derivative terms one needs to introduce a boundary counterterm. Therefore, in general the number of boundary counterterms coincides with the number of total derivatives one can construct using second derivatives of the longitudinal mode of the massive graviton.  
 
\vskip 0.15cm
\noindent
A theory of 5D massive gravity in $AdS_5$  with a 4D flat boundary has been proposed in \cite{Gabadadze:2017jom}  as a mechnism for raising the strong coupling scale of massive gravity to a significantly higher value. To avoid boundary ghosts, boundary counterterms were introduced in that framework in the decoupling limit, where the calculations of the strong scale were performed.~Our  goal here is to  present a fully diffeomorphism-invariant form of the boundary counterterms and study their consequences in Minkowski space (in the present work), as well as in $AdS$ space (in the sequel to this manuscript \cite {upcoming}). 
 
\vskip 0.15cm
\noindent
 The paper is organized as follows.~In the next section, we briefly review the dynamics of 5D general relativity in the presence of a spacetime boundary.~We present the fully diffeomorphism-invariant formulation of the theory and underline the role, played by the boundary bending mode in this formulation.~We then proceed, in Sec.~\ref{FP}, to the case of the Fierz-Pauli theory of a free massive graviton and show that this theory requires a novel boundary term (in addition to the Gibbons-Hawking term of GR) for consistency, once the background spacetime is endowed with a boundary.~We derive this term in two different ways: first by studying consistency of the boundary effective action in the 5D theory, and then by considering the Kaluza-Klein modes of a 6D massless graviton in the presence of a boundary, with one spatial dimension along the boundary compactified on a circle.~Sec.~\ref{w/obound} is mostly devoted to setting up notation and reviewing some standard results concerning 5D ghost-free massive gravity, that we find useful in the further parts of the paper.~In Sec.~\ref{secdl}, based on consistency of the decoupling limit, we generalize the boundary term of the linear FP theory to the full set of nonlinear and diffeomorphism-invariant boundary terms for ghost-free massive gravity.~Finally, in Sec.~\ref{conclusions} we conclude.

\section{General relativity with a boundary}
\noindent
Before we get to massive gravity, it will prove useful to briefly review the case of general relativity on a flat spacetime $\it{M}$ with a timelike codimension-1 boundary.~We will be particularly interested in the situation where $\it{M}$ is a five-dimensional manifold, parametrized by coordinates $X^M$ ($M=0,\dots,3,5$) and endowed with a metric $g_{MN}$, while its boundary $\p\textit{M}$ is given, at least in some coordinates (to be specified below), by the $X^5 \equiv z =0$ hypersurface.~We will further assume that the bulk metric is a small perturbation of Minkowski space: $g_{MN} = \eta_{MN} + h_{MN}$.

\begin{figure}[t]
\begin{center}
\includegraphics[width=7cm]{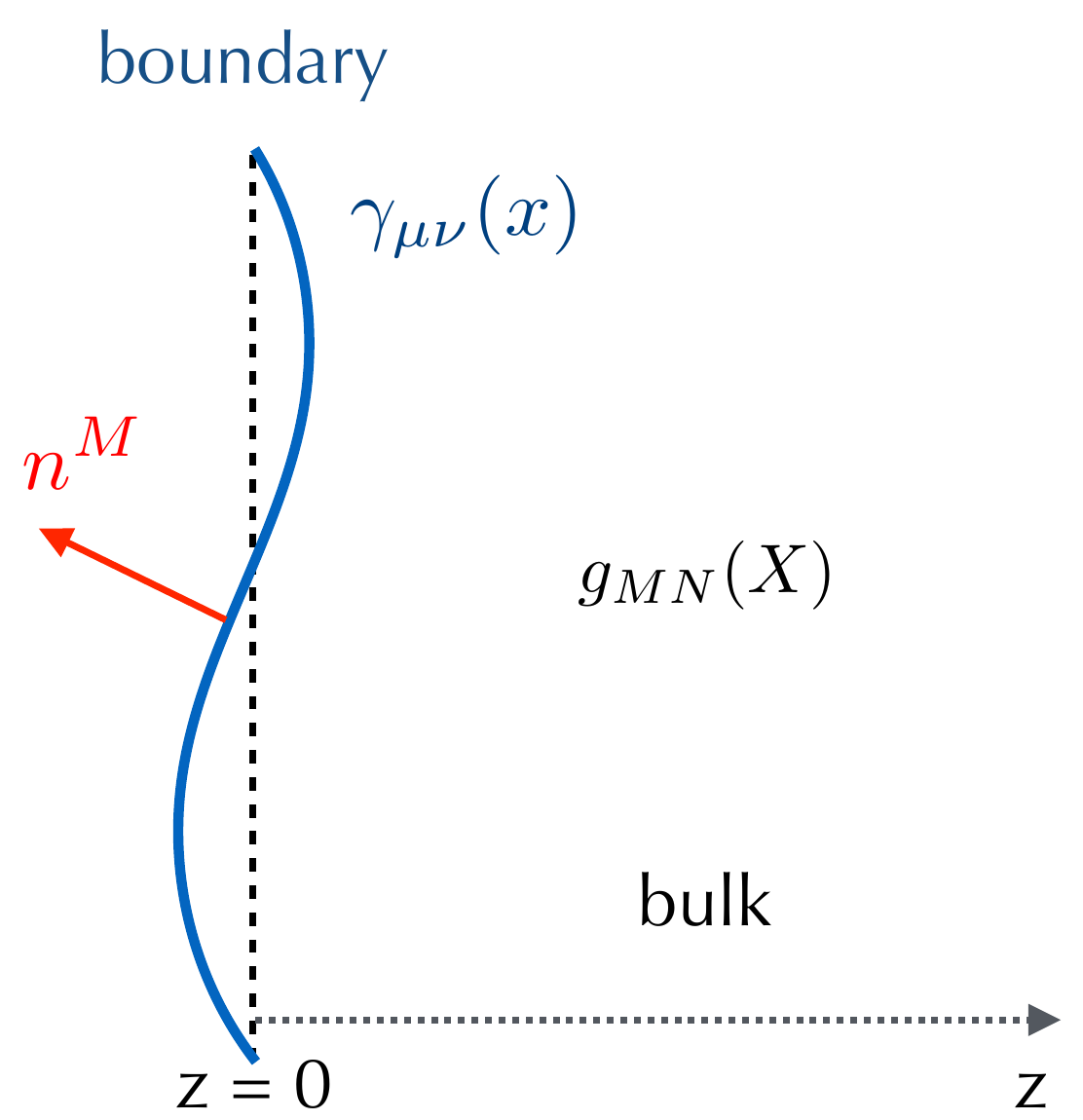} 
\caption{\label{fig1}~Embedding of the boundary in general coordinates.}
\end{center}
\end{figure}

\vskip 0.15cm
\noindent
 It is well-known that in order for the variational principle to be well-defined in GR, the Einstein-Hilbert action ought 
 to be amended by the Gibbons-Hawking boundary term:
\be
\label{nlact}
S_{\rm GR} = \frac{M_5^3}{2}\int_{\textit{M}} d^5 X~\sqrt{-g}~\upleft{5}R_5  - M_5^3 \int_{\p\textit{M}} d^4 x~\sqrt{-\gamma}~K~,
\ee
where $\upleft{5}R$ is the $5$-dimensional scalar curvature, $x^\mu$ ($\mu=0,\dots,3$) are the coordinates parametrizing the boundary manifold $\p\textit{M}$, $K$ denotes the (trace of the) extrinsic curvature associated with the embedding of  $\p\textit{M}$ in ambient spacetime, and $\gamma_{\mu\nu}$ is the induced metric on the boundary.~The 5D Planck mass is denoted by $M_5$.~The position of the boundary is described by five embedding functions $X^M(x)$, in terms of which the induced metric is given by the standard formula $\gamma_{\mu\nu}(x) = \p_\mu X^M\, \p_\nu X^N\, g_{MN}$.~Figure \ref{fig1} provides a cartoon of this setup.

\vskip 0.15cm
\noindent
There are two sets of gauge redundancies, associated with the action \eqref{nlact}: reparametrizations of the boundary (i.e.~4D diffs acting on the boundary coordinates $x^\mu$), and reparametrizations of the bulk (that is, 5D diffs acting on the bulk coordinates $X^M$). 
Bulk diffeomorphisms, $\delta X^M =\xi^M (X)$,\footnote{In a little more expanded form, what we mean by this notation is: $X^M =X'^M -\xi^M(X')$, or, to the leading order in the small diff parameter, $X'^M =X^M +\xi^M(X)$.~Therefore, if the boundary sits at $z=0$ in the `unprimed' coordinates, it will sit at $z' = \xi^z(X^\mu,0)$ in the `primed' coordinates.} act on the bulk metric $g_{MN}$ in the usual way
\be
\label{bulkdiffsbulkmetric}
\delta g_{MN}(X) = -\xi^K \p_K g_{MN}(X) - \p_M\xi^K g_{KN}(X) - \p_N\xi^K g_{KM}(X)\,,
\ee
while the induced metric on the boundary is invariant.~Under a change of the \textit{boundary coordinates}, on the other hand, the bulk metric and coordinates transform as scalars
\be
\label{boundarydiffs}
\delta x^\mu =  \zeta^\mu(x)\,, \quad \delta X^M =- \zeta^\mu\p_\mu X^M\,, \quad \delta g_{MN} = -\zeta^\mu \p_\mu X^K \p_K g_{MN}\,,
\ee
and the induced metric tranforms as a rank-2 tensor
\be
\label{bulkdiffsindmetric}
\delta \gamma_{\mu\nu} =- \zeta^\lambda\, \p_\lambda \gamma_{\mu\nu} - \p_\mu\zeta^\lambda \,\gamma_{\lambda\nu}-\p_\nu\zeta^\lambda \,\gamma_{\lambda\mu}\,.
\ee

\noindent
In practice, we will always use freedom to parametrize the boundary so that its coordinates are aligned with the first four of the bulk ones
\be
\label{ug}
X^\mu(x) = x^\mu\,, \quad X^5(x) = \chi(x)\,.
\ee
This is obviously not a complete gauge fixing yet. Bulk diffs change the boundary embedding functions, $
\delta X^\mu = \xi^\mu (x,\chi),~\delta X^5 = \xi^z(x,\chi),$
taking us away from the gauge \eqref{ug}; however, one can return to this gauge by performing a \textit{boundary} diffeomorphism of the form $\delta x^\mu = \xi^\mu \(x,\chi(x)\).$ This combined set of diffs make up the residual gauge freedom \textit{within} \eqref{ug}, under which the `boundary bending  mode' $\chi$ transforms as 
\be 
\label{deltapi}
\delta \chi = -\xi^\mu (x,\chi)\p_\mu \chi + \xi^z(x,\chi)\,,
\ee
while the transformation of the bulk metric is that given in eq.~\eqref{bulkdiffsbulkmetric}.~Furthermore, in the gauge at hand, the induced metric $\gamma_{\mu\nu}$ becomes
\be
\label{indmet}
\gamma_{\mu\nu}(x) = \p_\mu X^M\, \p_\nu X^N\, g_{MN} = g_{\mu\nu} + \p_\mu\chi g_{z\nu} + \p_\nu\chi g_{z\mu} + \p_\mu\chi \p_\nu\chi g_{zz}\,,
\ee
and obeys the standard transformation law \eqref{bulkdiffsindmetric} of a rank-2 tensor (with $\zeta^\mu(x) = \xi^\mu(x,\chi)$) under the residual diffs.

\vskip 0.15cm
\noindent
In the rest of this section, we will be interested in exploring the spectrum of small fluctuations about the bulk Minkowski vacuum.~To this end, we will need to perturb the action \eqref{nlact} to quadratic order in field fluctuations.~In the gauge \eqref{ug}, the relevant degrees of freedom are the bulk metric $h_{MN}$ and the beinding mode $\chi$, which lives on the boundary.~Furthermore, for the time being we will find it convenient to use the residual gauge freedom \eqref{deltapi} to fix 
\be
\chi(x) = 0~.
\ee
In this gauge, the boundary sits at $z =0$, which will significantly simplify the calculations.~We will later use the transformation properties of various fields under residual diffs to go back to a more general gauge, in which the boundary is bent.

\vskip 0.15cm
\noindent
It is a long, but straightforward exercise to expand the full action \eqref{nlact} to quadratic order in field fluctuations.\footnote{As a technical remark, we note that the presence of the boundary requires to keep careful track of all total $z$-derivatives that emerge from the bulk action.} The result is
\begin{align}
\label{quadact}
\begin{split}
\frac{2}{M_5^3}\, S_{\chi = 0} &= \int d^5 x~\frac{1}{8}h_{AB}\,\epsilon_{AMK}\epsilon_{BNL}\,\p_M\p_Nh_{KL} \\ 
&+\int_{z=0}\, d^4x\,\bigg[ \frac{1}{4} h_{\mu\nu}\p_z h_{\mu\nu}  -  \frac{1}{4} h_4\p_z h_4 + \frac{1}{2} h_4\p_\mu h_{z\mu} - \frac{1}{2} h_{\mu\nu}\p_\mu h_{z\nu}\bigg]\,,
\end{split}
\end{align}
where we have defined $h_4\equiv \eta^{\mu\nu} h_{\mu\nu}$ and the undisplayed indices on the two 5D fully antisymmetric epsilon symbols (our convention is $\epsilon_{01235} =1$) are understood as contracted among each other.\footnote{For example,~$\epsilon_{AMK}\epsilon_{BNL} \equiv \epsilon_{AMKPQ}\epsilon_{BNLPQ}\,.$}~All indices are contracted with the flat Minkowski metric and, for the sake of notational simplicity, we haven't raised/lowered them.
As a quick consistency check, we note that this action is invariant under the residual gauge transformations of the gauge \eqref{ug} that leave the boundary at $z=0$ (i.e.~those, compatible with the additional gauge fixing $\chi = 0$).~In other words, gauge transformations of the form \eqref{bulkdiffsbulkmetric} with the parameters $\xi^M(x,z)$ satisfying $\xi^\mu(x,0)\neq 0$ and $\xi^z(x,0)= 0$ are an exact symmetry of \eqref{quadact}.

\vskip 0.15cm
\noindent
To arrive at a fully diff-invariant description of the theory, we will now explicitly restore the bending mode $\chi$ in the action by performing a diffeomophism of the form \eqref{deltapi}. This amounts to redefining $z = z' - \xi^z(x,z')$, with $\xi^z\big | =\chi(x)$ (here and in what follows, a vertical stroke without a subscript will denote evaluation on the boundary), and substituting for the metric $h_{MN}(x,z) = h'_{MN}(x,z') + \p_M\xi_N(x,z') + \p_N\xi_M(x,z')$.~The boundary is located at $z' = \chi(x)$ in the new coordinates.~Performing the above substitutions (and omitting primes on new fields and coordinates for notational simplicity) turns \eqref{quadact} into 
\begin{align}
\label{quadactwithbb}
\frac{2}{M_5^3}\, S_{\rm GR}\big[h_{MN},\chi\big]=
\frac{2}{M_5^3}\, S_{\chi = 0}\big[h_{MN}] -\int_{z=\chi} d^4 x~ \chi\,(\p_\mu\p_\nu h_{\mu\nu} - \p^2 h_4)\,,
\end{align}
where $S_{\chi = 0}$ is given in eq.~\eqref{quadact}, and we have defined $\p^2\equiv \eta^{\mu\nu}\p_\mu\p_\nu$.~Although the boundary is now \textit{not} at $z=0$, at the quadratic order in field fluctuations we can still integrate the second term in \eqref{quadactwithbb} over the $z=0$ hypersurface, since the difference between this and integrating over $z=\chi(x)$ is at least cubic in the fields.~One can check explicitly that \eqref{quadactwithbb} is invariant under the full set of 5D diffeomorphisms of the form
\be
\label{diffbdy}
\delta h_{MN} = \p_M \xi_N +\p_N \xi_M\,, \qquad \delta \chi = -\xi^z\,,
\ee
with \textit{arbitrary} $\xi^{M}(x,z)\,.$
\vskip 0.15cm
\noindent
Eqs \eqref{quadact} and \eqref{quadactwithbb} provide the starting point for the discussion of (Fierz-Pauli) massive gravity in the presence of spacetime boundaries, which we will turn to next. 

 \section{A boundary term for the Fierz-Pauli theory}
\label{FP}
\noindent
Having obtained the fully diffeomorphism-invariant formulation of GR on on a spacetime with a boundary, we wish to explore possible ways of generalizing it to the case of massive gravity.

\vskip 0.15cm
\noindent
As a matter of fact, it is instructive to first recall how things work \textit{in the absence} of boundaries. 
The unique action for a free, massive spin-2 particle on 5D Minkowski space is \cite{Fierz:1939ix}
\be
S_{\rm FP} = S_{\rm EH} - \frac{1}{8}\,M_5^3 m^2\int d^5 x\, (h^{AB}h_{AB}-h_5^2)\,,
\ee
where $S_{\rm EH}$ denotes (the quadratic piece) of the Einstein-Hilbert lagrangian and $m^2$ is the particle's squared mass.~The massive theory, in contrast to the massless one, has no redundancy in terms of the field $h_{AB}$ alone.

\vskip 0.15cm
\noindent
We will follow the same logic in the presence of a spacetime boundary at $z=0$ and define massive gravity as a deformation of the massless theory, where the full 5D diff invariance is recovered by formally setting the mass parameter to zero.~As we have seen in the previous section, 5D diff invariance necessarily implies the presence of the bending mode $\chi$, which lives on the boundary and nonlinearly realizes local translations along the $z$ direction.~We will therefore require the putative massive gravity action to reduce to \eqref {quadactwithbb} for $m^2 = 0$.

\vskip 0.15cm
\noindent
Running a bit ahead of the discussion, we will state here the expression for the correct action of a massive spin-2 particle in the presence of spacetime boundaries:
\be
\label{quadactwithbb1}
\begin{split}
\frac{2}{M_5^3}\, S &= \int d^5 x~\bigg[\frac{1}{8}h_{PQ}\,\varepsilon_{PMK}\varepsilon_{QNL}\,\p_M\p_Nh_{KL} -\frac{m^2}{4}\, (h^{MN}h_{MN}-h_5^2)\bigg]\\
&+\int_{z=0}\, d^4x\,\bigg[ \frac{1}{4} h_{\mu\nu}\p_z h_{\mu\nu}  -  \frac{1}{4} h_4\p_z h_4 + \frac{1}{2} h_4\p_\mu h_{z\mu} - \frac{1}{2} h_{\mu\nu}\p_\mu h_{z\nu}\\ 
&- \chi\,(\p_\mu\p_\nu h_{\mu\nu} - \p^2 h_4)- m^2 \chi h_4\bigg].
\end{split}
\ee
A few remarks are in order.~First, this indeed reproduces the fully diff-invariant action \eqref{quadactwithbb} of general relativity once the parameter $m^2$ is set to zero.  
Second, the first line in \eqref{quadactwithbb1} is completely fixed by consistency of the bulk dynamics to be that of Fierz and Pauli (locality in the extra dimension implies that the presence of the boundary can not influence the bulk part of the action).~Furthermore, generically one expects a modification of the \textit{boundary} action, proportional to the mass parameter $m^2$ as well.~This is given by the extra term
\be 
\label{fpbdy}
\frac{2}{M_5^3}\, \mathcal{L}_{\rm FP}^{\rm bdy} = - m^2\chi h_4 \big|
 \ee
we've added on the last line.~Deriving this boundary operator  and exploring its consequences will be the main focus of the rest of this section. 

\subsection{The boundary effective action}
\noindent
In order to better understand the meaning of the boundary term \eqref{fpbdy} for Fierz-Pauli massive gravity, one can compute the \textit{boundary effective action} that results from integrating out the bulk degrees of freedom (we will use the method of  ref.~\cite{Luty:2003vm} that constructed  the boundary effective action in the DGP model \cite{Dvali:2000hr}).~The bulk equations of motion that follow from varying the action \eqref{quadactwithbb} with respect to $h_{MN}$ are the usual Fierz-Pauli equations, which in the absence of bulk sources imply 
\be
\label{bulkeqs}
\p_M h_{MN} = h_5 = 0~, \quad (\Box - m^2 )h_{MN} = 0\,,
\ee
where we have defined $\Box \equiv \eta^{MN}\p_M\p_N$ and $h_5 = \eta^{MN} h_{MN}$.~The solution of the last of these equations that obeys the physical, `outgoing wave' boundary conditions at infinity is\footnote{In Euclidean space, this solution corresponds to the fields decaying at infinity.}
\be
h_{MN}(x,z) = e^{-\Delta z}\,\tilde h_{MN} (x)\,,\qquad \Delta \equiv \sqrt{-\p^2+m^2}\,,
\ee
where the boundary values of the bulk metric, $\tilde h_{AB} = h_{AB}\big |$, are four-dimensional fields with respect to which the boundary effective action will be later varied. 
The first two of eqs.~\eqref{bulkeqs}, which in 4D notation give 4+1+1 conditions, allow to express $\tilde h_{\mu z}$ and $\tilde h_{zz}$ in terms of $\tilde h_{\mu\nu}$, as well as impose a further equation on $\tilde h_{\mu\nu}$ itself. On-shell, this amounts to the following relations
\be
\label{constraints}
\tilde h_{zz} = -\tilde h_4~, \qquad \tilde h_{\mu z} = \frac{1}{\Delta} \p_\nu \tilde h_{\nu\mu}~,\qquad  \p_\mu\p_\nu \tilde h_{\mu\nu} -\p^2 \tilde h + m^2 \tilde h = 0\,.
\ee
The last equation here is of particular importance: it exactly coincides with the  equation one obtains upon varying the boundary action in \eqref{quadactwithbb1} with respect to the bending mode $\chi$.~Notice that the boundary term \eqref{fpbdy} plays a crucial role: had we added it with a different coefficient---or, for that matter, omitted it altogether---the equation of motion of $\chi$ would be inconsistent with the (on-shell) constraints coming from the bulk physics.~We stress once again that the bulk equations of motion/constraints can not be modified --- they are fixed to be those of Fierz and Pauli by the absence of ghosts and spacetime boundaries can not influence this fact due to locality in the extra dimension. 

\vskip 0.15cm
\noindent
Using eqs.~\eqref{constraints} in \eqref{quadactwithbb1} yields the following effective action for the boundary degrees of freedom:
\begin{align}
\frac{2}{M_5^3} \,S_{\rm bdy}  =  \frac{1}{4}  \int d^4x~\bigg[\tilde h_{\mu\nu}\frac{1}{\Delta}\varepsilon_{\mu\alpha\rho}\varepsilon_{\nu\beta\sigma}\p_\alpha\p_\beta \tilde h_{\rho\sigma}-m^2\(\tilde h_{\mu\nu}\frac{1}{\Delta}\tilde h_{\mu\nu} - \tilde h \frac{1}{\Delta}\tilde h\)   \bigg].
\end{align}
(The boundary bending mode $\chi$, being a Lagrange multiplier, falls out from the on-shell action.)~Up to the (operator) factors of $\Delta^{-1}$, this is just the standard EH kinetic term plus the Fierz-Pauli mass.~Nonlocality of the 4D effective description implied by the form of the 2-point function, $\langle h h \rangle \sim \Delta/\p^2$, has a clear physical interpretation: it corresponds to the continuum of massive Kaluza-Klein modes, that make the theory truly five-dimensional.

\vskip 0.15cm
\noindent
In the next subsection, we will derive the boundary term \eqref{fpbdy} from a different perspective. 

\subsection{The boundary term for the FP theory from 6D general relativity}
\noindent
Consider general relativity in six spacetime dimensions, with the extra two (spatial) dimensions denoted by $X^5 \equiv z$ and $X^6\equiv y$.~In this subsection (and only in this subsection), capital Latin indices will run over $M,N = 0,1,2,3,5,6$, lower-case Latin indices will run over $m,n = 0,1,2,3,6$, and Greek indices will correspond to the standard 4D coordinates, $\mu,\nu = 0,1,2,3$.~Like before, we will assume that the background manifold is endowed with a boundary at $z=0$.~Moreover, the sixth dimension will be assumed to be compactified on a circle, $y \sim y + 2\pi L$\,.~From the 5D point of view, the spectrum thus contains the usual 5D graviton, vector and scalar zero modes, as well as the massive Kaluza-Klein modes of the 6D graviton, that have definite momenta in the $y$ direction.~Furthermore, due to the boundary at $z=0$, in a generic gauge there will also be the $5D$ bending (zero-)mode $\chi^{(0)}(x^\mu)$, along with its KK modes $\chi^{(n)}(x^\mu)$, localized on the boundary.

\vskip 0.15cm
\noindent
The graviton's KK modes are massive 5D spin-2 particles in a flat spacetime with a boundary, described by the Fierz-Pauli action in the bulk.~The full theory, being just 6D general relativity (with a boundary and one compactified direction), is obviously consistent, and thus the theory of the graviton's KK modes that results from compactification should be consistent too.~According to our discussion above, any consistent theory of a massive spin-2 particle should include the boundary term \eqref{fpbdy} in the action.~We will now show how this term arises in the 6D setup at hand.

\vskip 0.15cm
\noindent
The 6D version of the quadratic action \eqref{quadactwithbb1} for general relativity with a spacetime boundary at $z=0$ reads
\begin{align}
\label{quadact6d}
\begin{split}
\frac{2}{M_6^4}\, S_{\text{6D}} &= \int d^6 x~\frac{1}{24}\,h_{AB}\,\epsilon_{AMK}\epsilon_{BNL}\,\p_M\p_Nh_{KL} \\ 
&+\int_{z=0}\, d^5x\,\bigg[ \frac{1}{4} h_{mn}\p_z h_{mn}  -  \frac{1}{4} (h_4 + h_{66})\p_z (h_4 + h_{66}) + \frac{1}{2} (h_4 + h_{66})\p_m h_{zm}\\ & - \frac{1}{2} h_{mn}\p_m h_{zn} - \chi \big[\p_m\p_n h_{mn} - \eta^{mn}\p_m\p_n (h_4 + h_{66})\big]\bigg]\,,
\end{split}
\end{align}
where $h_4 = \eta^{\mu\nu} h_{\mu\nu}$ as before, and we have denoted the 6D Planck mass by $M_6$.~Upon compactifying the $y$ direction, the fields are decomposed into their respective KK modes as follows
\be
\label{kkmodes}
h_{MN} = \frac{1}{\sqrt{L}}\sum^{+\infty}_{n = -\infty} h^{(n)}_{MN}(x^\mu, z)\, e^{iny/L}\,, \qquad \chi =\frac{1}{\sqrt{L}} \sum^{+\infty}_{n = -\infty} \chi^{(n)}(x^\mu)\, e^{iny/L}\,,
\ee
where $h^{(-n)}_{MN} = h^{(n)\,*}_{MN}$ and analogously for $\chi$, due to reality of the position-space fields.~In what follows, it will be convenient to work in the gauge
\be
\label{unitgaug}
h^{(n)}_{66} = h^{(n)}_{6 z} = h^{(n)}_{6\mu} = 0\,,
\ee
which is always possible to fix for the graviton's KK modes (i.e.~for modes with $n$ different from zero).~The bulk action for these modes, obtained by substituting the decomposition \eqref{kkmodes} into the bulk part of \eqref{quadact6d}, is then the Fierz-Pauli action in the first line of eq.~\eqref{quadactwithbb1}, with the KK masses given by 
\be
m_n =\frac{n}{L}\,.  
\ee
Let us now focus on the \textit{boundary} part of the action.~Again, by inserting the KK decomposition \eqref{kkmodes} into the second and the third lines of the 6D action \eqref{quadact6d} fixed to the gauge \eqref{unitgaug}, we find that the resulting theory at $z=0$ precisely reproduces the boundary action of Fierz-Pauli massive gravity, eq.~\eqref{quadactwithbb1}, at each KK level $n\neq 0$ (we identify $M_5^3 = M_6^4 L$).~Crucially, this boundary action includes the terms of the form \eqref{fpbdy} for every KK mode:
\be 
\label{fpbdy1}
\frac{2 }{M_5^3}\, \mathcal{L}^{(n)}_{\rm bdy} = - m_n^2\,\(\chi^{(n)\, *} h^{(n)}_4 + \rm{h.c.}\)\big|
 \ee
which come from the last term in \eqref{quadact6d}: 
\begin{align}
\begin{split}
2\, \mathcal{L}^{(n)}_{\rm bdy} = M_6^4\,\int _{\p\it{M}}d^4 x\, dy\,\,\chi\, \eta^{mn}\p_m\p_n h_4 &\supset M_5^3\,\sum_n \int _{\p\it{M}}d^4 x\, \chi^{(n)\,*}\, \p_6^2 h^{(n)}_4 + \rm{h.c.}\\&= -M_5^3\,\sum_n\int _{\p\it{M}}d^4 x\, \,m_n^2\,\chi^{(n)\,*}\, h^{(n)}_4 + \rm{h.c.}
\end{split}
\end{align}
These boundary terms have exactly the right coefficients to make the equations of motion for (KK modes of) the boundary bending mode consistent with the on-shell Fierz-Pauli equations/constraints in the bulk -- something we discussed at length in the previous subsection. 

\vskip 0.15cm
\noindent
Let us reiterate the results so far: we have shown that the boundary term \eqref{fpbdy} is required by consistency of Fierz-Pauli massive gravity on background manifolds with a boundary.~Without this term, the (on-shell) bulk constraints that the graviton field $h_{AB}$ satisfies would be inconsistent with the equation of motion for the boundary bending mode $\chi$.~We have also shown that \eqref{fpbdy} arises in 6D \textit{general relativity} with a spacetime boundary, upon compactifying one of the directions along the boundary on a circle.~In this setup, the 5D effective action contains boundary terms of precisely the form \eqref{fpbdy} for every massive 5D KK mode of the 6D graviton.~In Sec.~\ref{secdl}, we will unveil another face of the boundary term  \eqref{fpbdy} that becomes visible in a certain short-distance/high-energy limit of the theory, reviewed in the next section.

\section{Nonlinear massive gravity and the decoupling limit}
\label{w/obound}
\noindent
Mostly for the purposes of setting up notation and presenting results that we will find useful in later sections, we will summarize here the dynamics of massive spin-2 fields on $5$-dimensional (infinite) Minkowski spacetime. The discussion in this section is standard.

\vskip 0.15cm
\noindent
The unique  ghost-free non-linear extension of Fierz-Pauli theory of a free massive graviton in 4D has been found in refs.~\cite{deRham:2010ik, deRham:2010kj}. Its counterparts  for an arbitrary dimension $D\geq 3$ are straightforward to write.~In a diffeomorphism-invariant formulation, 5D massive general relativity features 5 auxiliary scalar fields, $\Phi^I$ $(I = 0,\dots,3, 5)$, in addition to the 5-dimensional metric. Explicitly, the action reads
\be
\label{nobound}
\begin{split}
S &=\frac{M_5^{3}}{2}\int d^5 X\,\sqrt{-g}\, \bigg[\upleft{5}R -\frac{m^2}{4} \sum_{n=2}^5\alpha_n U_n(\mathbb{K})\bigg]~,
\end{split}
\ee
where $\alpha_n$ are constant parameters, and $m$ is the graviton's mass (which fixes $\alpha_2 = 2/3$ in five dimensions).~Furthermore, the `potential' terms $U_n$ can be written with the help of the 5D totally antisymmetric symbol $\epsilon$ as follows\footnote{The usual cosmological constant corresponds to $n=0$, while the term linear in $\mathbb{K}^M_N$ (corresponding to $n=1$) leads to a tadpole on the Minkowski background and thus obstructs having a Poincar\'e-invariant vacuum. We will discard these terms in the rest of this paper.} 
\be
\label{Vs}
U_n =\epsilon_{M_1 \dots M_n M_{n+1}\dots M_5} \epsilon^{N_1 \dots N_n N_{n+1}\dots N_5}~\mathbb{K}^{M_1}_{N_1}\dots \mathbb{K}^{M_n}_{N_n}~\delta^{M_{n+1}}_{N_{n+1}}\dots \delta^{M_5}_{N_5}\equiv \epsilon_5 \epsilon_5\times \mathbb{K}^n \times \mathbb{1}^{5-n}~,
\ee
where the matrix $\mathbb{K}$ is defined in terms of the auxiliary scalars and the metric in the following way
\be
\label{K}
\mathbb{K}^M_N = \delta^M_N- \(g^{MK}\mathbb{f}_{KN}\)^{1/2}\,,\qquad \mathbb{f}_{MN} = \p_M\Phi^I \p_N\Phi^J\, \eta_{IJ}\,.
\ee
Here $\mathbb{f}_{IJ}$ is a flat auxiliary metric, related to the Minkowski one by a coordinate transformation. One can further generalize the theory by defining it with a \textit{curved} $\mathbb{f}_{IJ}$ \cite{Hassan:2011tf}, or even by promoting $\mathbb{f}_{IJ}$ to a full-fledged dynamical tensor field, which would define a bigravity theory \cite{Hassan:2011zd}.~The second equality in \eqref{Vs} defines notational shortcut, that we will often use in the rest of this paper.~We will assume that the metric is coupled minimally to matter, as it is in general relativity.~Finally, we note that the action \eqref{nobound} is invariant under the \textit{internal} $SO(4,1)$ rotations, acting on the auxiliary scalars' flavor index
\be
\label{internagroup}
\Phi^I \to \Lambda^I_{~J}\,\Phi^J\,,
\ee
where $\Lambda$ denotes a general 5D Lorentz transformation matrix.

\vskip 0.15cm
\noindent
The equations of motion that follow from varying the action \eqref{nobound} admit a flat-space solution with the following expectation values
\be
\label{scalavevs}
\langle g_{MN} \rangle = \eta_{MN}, \qquad \langle \Phi^I \rangle = \delta^I_M X^M.
\ee
On this background, the scalars' internal indices mix with the spacetime ones, so  from now on, we will often not make distinction between these.~One can use diffeomorphism invariance of the full action \eqref{nobound} to fix \textit{unitary gauge}, in which the five scalars are frozen to their background values, $\Phi^M  = X^M$.~In this gauge, \eqref{nobound} describes a Lorentz-invariant theory of the spin-2 field $h_{MN}$ alone.~Generic mass and potential terms for the graviton lead to loss of all (Hamiltonian and momentum) constraints of general relativity. As a result, a generic theory of massive gravity propagates one extra degree of freedom in addition to the nine, required by the representation theory of the 5D Poincar\'e group.~This extra dof is necessarily a (Boulware-Deser) ghost \cite{Boulware:1973my}.~In contrast, the special structure of the action \eqref{nobound}-\eqref{K} guarantees that one combination of the Hamiltonian and momentum constraints persists even for $m^2\neq 0$, projecting out the unwelcome sixth degree of freedom and rendering the theory ghost-free \cite{deRham:2010ik,deRham:2010kj}. 

\vskip 0.15cm
\noindent
In more physical terms, the potential issues associated with the Boulware-Deser ghost, and cure thereof, can be grasped by considering the spin-2 analog \cite{ArkaniHamed:2002sp} of the \textit{Goldstone equivalence limit} in massive vector theories \cite{Cornwall:1974km}.~It is customary to refer  to this as the \textit{decoupling limit}.~The virtue of this limit is that it distills the high-energy dynamics of the various physical polarizations of the spin-2 multiplet.\footnote{Corresponding to a \textit{ultraviolet} instability, the BD ghost is clearly visible in this limit.}~In particular, away from unitary gauge and at high energies, the helicity-0 mode $\Pi$ and helicity-1 mode $A^M$ of the massive graviton are embedded in the metric and the auxiliary scalars in the following way
\be
\label{h0}
h_{MN} =\frac{1}{M_5^{3/2}} \(\hat h_{MN} + \frac{2}{3}\Pi\eta_{MN}\)\,,\qquad \Phi^M = X^M +\frac{V^M}{M_5^{3/2} m} \,,
\ee
where $V^M$ is further decomposed as 
\be
\label{h1}
V^M = A^M - \frac{\p^M\Pi}{m}\,.
\ee
Here $\hat h_{MN}$ describes the graviton's helicity-2 polarization at short distances/high energies.~For further reference, we also note an important property of the tensor $\mathbb{K}^M_N$\,, whereby it takes on a simple form upon substituting $h_{MN} = A^M =0$:
\be
\label{propofk}
\mathbb{K}^M_N\big|_{h_{MN}=A^M=0} = \frac{\p^M\p_N\Pi}{\Lambda_{7/2}^{7/2}}\,.
\ee
In writing this formula, we have defined $\Lambda_{7/2}\equiv \big(M_5^{3/2} m^2\big)^{2/7}$ and used the vertical stroke with a subscript to denote conditional evaluation (not to be confused with a stroke without a subscript, which  denotes evaluation on the boundary). Importantly, eq.~\eqref{propofk} holds at the full non-linear level.

\vskip 0.15cm
\noindent
The decoupling limit (DL) is defined as a double scaling limit, in which
\be
\label{declim}
M_5 \to \infty~, \qquad m\to 0~, \qquad \Lambda_{7/2} = \text{finite}.
\ee
Making use of eqs.~\eqref{K},~\eqref{h0} and \eqref{propofk} and expanding the action of massive GR \eqref{nobound} in powers of the various fields over the Minkowski vacuum, one finds that the only singular terms in the  limit \eqref{declim} are \cite{ArkaniHamed:2002sp}
\be
\label{piaction}
S_{\rm sing} = -\frac{1}{8} \, \int d^5 x ~ \frac{\Lambda_{7/2}^7}{m^2}\,  \sum_{n=2}^5 ~\frac{\alpha_n}{\Lambda_{7/2}^{7n/2}}~\epsilon_5\epsilon_5\times \(\p^2_{\mathbb{5}}\Pi\)^n \times \mathbb{1}^{5-n}~.
\ee
In writing this expression,~we have used the matrix notation defined in eq.~\eqref{Vs} and schematically denoted a $D$-dimensional partial derivative by $\p_\mathbb{D}$, so that $\p^2_{\mathbb{5}}\Pi$ is short for the matrix $\p^M\p_N\Pi$. 

\vskip 0.15cm
\noindent
The specific structure of the terms in \eqref{piaction} directly follows from the form of the potential \eqref{nobound} and eq.~\eqref{propofk}.
All terms in \eqref{piaction} apart from these are regular in the decoupling limit. What allows for this limit to be well-defined is then the fact that the $(\p_{\mathbb{5}}^2\Pi)^n$ terms, while formally divergent, combine into total derivatives and are thus immaterial as long as the background spacetime has no boundaries and the fields decay at infinity fast enough. This in a sense is the defining feature of the particular structure of mass and potential terms given by eqs.~\eqref{Vs}, \eqref{K} and \eqref{propofk}: this structure renders the classical theory ghost-free, while the quantum theory becomes endowed with a relatively high cutoff $\Lambda_{7/2}$.~In a \textit{generic} theory with massive spin-2 fields, on the other hand, the operators of the form $(\p_{\mathbb{5}}^2\Pi)^n$ do not gather into total derivatives and do genuinely affect the dynamics. This leads to a ghost instability in the classical theory, and to a low cutoff $\Lambda_{11/2} = (M_5^{3/2} m^4)^{2/11}$ once the theory is viewed as a low-energy quantum EFT. 

\vskip 0.15cm
\noindent
The above discussion immediately suggests that things should change in the presence of spacetime boundaries.~In that case, the higher-derivative self-interactions of the helicity-0 field $\Pi$ \textit{are not} entirely eliminated from the dynamics, but do affect physics at the boundary, leading to a higher-derivative boundary action and the associated ghost instabilities.~We will thus find that consistency of the decoupling limit \eqref{declim}, and for that matter -- of the full theory, will require defining the \textit{boundary action} of massive GR such that the effects of the bulk operators of the form $(\p_{\mathbb{5}}^2\Pi)^n$ are cancelled.~Understanding how this works in detail will be our main task in the rest of this paper. 

 \section{Nonlinear massive gravity with  a boundary}
\label{secdl}
\noindent
In this section, we will address the question of how to define the decoupling limit of massive gravity in the presence of spacetime boundaries.~As discussed above, the decoupling limit is convenient in that it makes the dynamics of various helicity modes of the massive spin-2 particle transparent at energies, higher then the particle's mass. Strictly speaking, spacetime boundaries break translation invariance in one or more directions, and it may seem that working in terms of the representations of the Poincar\'e group---the helicity modes---puts one on shaky grounds.~This logic is misleading, however.~First of all, we have seen that there is no obstruction whatsoever to restoring the local translation invariance of the theory by means of introducing new fields (in our case, the boundary bending mode $\chi$). And second: as already emphasized above, it is physically clear that the presence of boundaries can not influence the bulk physics simply by locality.~The (high-energy) bulk dynamics should thus admit a perfectly adequate description in terms of the modes of definite helicity. 

\subsection{The boundary term for the FP theory: a third derivation}
\label{secdl1}
\noindent
With the above considerations in mind, we procceed by first defining the decoupling limit for our \textit{quadratic} massive gravity action \eqref{quadactwithbb1} that includes the novel boundary term \eqref{fpbdy}.~To this end, we restore the diffeomorphism invariance, `broken' by the presence of the bulk and boundary terms, proportional to the mass parameter $m^2$.~This is achieved by introducing the \stu field $V^M$\, along the lines of eq.~\eqref{h0} of the previous section
\be
\label{stu1}
h_{MN} \to \frac{1}{M_5^{3/2}}\,\bigg [ h_{MN} - \frac{1}{ m} \(\p_M V_N + \p_N V_M\)\bigg ]\,, \quad \chi \to   \frac{1}{M_5^{3/2}}\,\(\chi + \frac{1}{m} V^z\big |\)\,.
\ee
Notice that we have also canonically normalized the various fields along the way, but have kept the old notation for the new fields; we will be careful to comment whenever this may lead to confusion.~Eq.~\eqref{stu1} resembles a broken linearized diff of the form \eqref{diffbdy}.~As remarked in Section~\ref{w/obound}, at high energies the vector $V^M$ further decomposes into the helicity-1 ($A^M$) and helicity-0 ($\Pi$) polarizations of the massive graviton as in eq.~\eqref{h1}.~The theory of massive gravity \eqref{quadactwithbb1}, defined in terms of the new fields  is thus invariant under the full set of linearized 5D diffeomorphisms 
\begin{align}
\delta h_{MN} =  \p_M \xi_N + \p_N \xi_M\,, \quad \delta \chi = -\xi_z\big |\,,\quad \delta A_M = m\,\xi_M\,, \quad \delta\Pi = 0\,,
\end{align}
as well as an additional local $U(1)$ group
\begin{align}
\delta h_{MN} = 0\,, \quad \delta \chi = 0\,,\quad \delta A_M = \p_M\alpha\,, \quad \delta\Pi = m\alpha\, .
\end{align}
Notice that the scaling of $h_{MN},\,A_M$ and $\Pi$ with $m$ is fully fixed by the well-known dynamics of Fierz-Pauli massive gravity in the bulk: in order to continuously reproduce the correct number of the high-energy degrees of freedom, these (canonically normalized) fields ought to scale as $m^0$.~The presence of the spacetime boundary can not change this fact.

\vskip 0.15cm
\noindent
We'd now like to zoom onto distcance scales, much shorter than the graviton's Compton wavelength, which operationally amounts to sending the paramter $m^2$ to zero, while the rest of the parameters scale as in eq.~\eqref{declim}.~Substituting the field decomposition \eqref{stu1} and \eqref{h1} into the massive gravity action \eqref{quadactwithbb1} and
focussing for the time being on the part that features $V_M$ alone, we have
\begin{align}
\label{vlag}
2S_V = \int\,d^5x\,\bigg[-\frac{1}{4}F^{MN}F_{MN} - \p_M\(V_N\,\p_N V_M - V_M\, \p_N V_N\) \bigg ]+\int_{z=0}d^4x ~2V_z\, \p_\mu V_\mu~,
\end{align}
where $F \equiv dV = dA$ is the usual field strength for the vector $V$.~The first integral in this expression comes from the buk mass in \eqref{quadactwithbb1}, while the second, boundary integral comes from our new boundary term \eqref{fpbdy}.~Now, the vector $V^M$ is further decomposed into the helicity-1 (vector) and helicity-0 (scalar) modes as in \eqref{h1}. The scalar mode obviously falls out from the first, Maxwell term in the bulk integral.~It is however present in the second, total derivative contribution to this integral, as well as in the last, boundary term in \eqref{vlag}.~Given that the canonically normalized $\Pi$ does not scale with the graviton mass in \eqref{h1}, these terms separately diverge as $m^{-2}$ (see the discussion of the bulk terms around eq. \eqref{piaction} of Section \ref{w/obound}).~This divergence potentially obstructs validity of the $\Lambda_{7/2}$ decoupling limit.~However, one can check that while divergent separately, the unwelcome bulk and boundary terms `magically' cancel against each other, once the bulk total derivative is rewritten as a boundary integral.~Again, the presence of the boundary term \eqref{fpbdy} has played a crucial role in this discussion. Without this term, one would not be able to define the standard $\Lambda_{7/2}$-decoupling limit of massive gravity on spacetimes with boundaries: the unique prescription that continuoulsy reproduces the high-energy degrees of freedom of the spin-2 particle in the bulk would be ill-defined on the boundary.

\vskip 0.15cm
\noindent
Having taken care of the terms that diverge like $m^{-2}$ in the DL, we now turn to exploring those that scale like $m^0$.~Apart from the Maxwell term for $A^M$ that we have already discussed above, these terms feature the tensor and scalar modes $h_{MN}$ and $\Pi$, as well as the boundary bending mode $\chi$.~Moreover, it is a well-known fact that in the given basis of fields, $\Pi$ has no kinetic term and its bulk dynamics comes entirely from the kinetic mixing with the tensor mode, $S_{\rm mix}\sim \int \,d^5x\, h_{MN} (\p_M\p_N\Pi - \eta_{MN}\Box\Pi)$.~In order to diagonalize the bulk kinetic Lagrangian, one then performs a conformal transformation on $h_{MN}$ \cite{ArkaniHamed:2002sp}
\be
\label{conftr}
h_{MN} = \hat h_{MN} +\frac{2}{3}\Pi\eta_{MN}\,.
\ee
In the new (and final) field basis, the \textit{bulk} action becomes
\begin{align}
\label{bulkdl}
\frac{2}{M_5^3}\, S_{\rm bulk} &= \int d^5 x~\bigg[\frac{1}{8}\hat h_{PQ}\,\varepsilon_{PMK}\varepsilon_{QNL}\,\p_M\p_N \hat h_{KL} \nn \\&+\frac{1}{2}\p_M \big(\Pi\p_N \hat h_{MN} - \p_N\Pi \hat h_{MN}+\p_M\Pi \hat h -\Pi \p_M \hat h\big)+\frac{4}{3}\Pi\Box\Pi\bigg]\,.
\end{align}
Again, there is a total derivative involved, which originates from manipulating terms that come from the graviton mass terms in the bulk.~An important point is that in our case of a background spacetime with a boundary, the conformal transformation \eqref{conftr} affects not only the bulk part of the action in \eqref{quadactwithbb1}, but the boundary one as well.~The transition to the new field basis generates a whole host of new terms \textit{on the boundary} that, as one can check by a direct calculation, non-trivially cancel against the pre-existing boundary terms as well as the bulk total derivative in \eqref{bulkdl} (once it is written as a boundary term).~Again, the contribution of the massive gravity boundary operator \eqref{fpbdy} that survives in the decoupling limit, 
\be
\lim_{m\to 0}\frac{2}{M_5^3}\,\mathcal{L}^{\rm FP}_{\rm bdy} = \(-2\chi \p^2\Pi + h\p_z\Pi \)\big |= \(-2\chi \p^2\Pi +\hat h\p_z\Pi +\frac{8}{3}\Pi\p_z\Pi\)\bigg |
\ee
is absolutely crucial for these cancellations to occur.~Once the dust settles, we find that the single conformal transformation \eqref{conftr} that diagonalizes the bulk action, also `magically' diagonalizes the boundary one,\footnote{Again, without adding the boundary term \eqref{fpbdy}, this would not be true.} so that the complete quadratic decoupling limit action becomes
\be
\label{dllag}
S^{\rm DL}_{\rm FP} = S_{GR}\big[\hat h_{MN}, \chi\big] + \int\,d^5x\, \bigg[-\frac{1}{4}F^{MN}F_{MN} +\frac{4}{3}\Pi\Box\Pi\bigg] + \int_{z=0}\,d^4 x\, \frac{4}{3}\,\Pi\p_z\Pi\,,
\ee
where $S_{\rm GR}$ is the action \eqref{quadactwithbb} of general relativity with the bending mode in place.~Apart from the standard boundary term in $S_{GR}$, the only remnant of the above-mentioned cancellations is the last term in \eqref{dllag}, that only involves the helicity-0 mode of the graviton.~Upon rewriting this term as a bulk total derivative, and combining it with the existing bulk kinetic term of $\Pi$, we get for the helicity-0 action:
\be
S^{\rm DL}_{\Pi} = \int\,d^5x\,-\frac{4}{3}\p_M\Pi\p_M\Pi\,,
\ee
so that the graviton's scalar mode enters the decouling limit action with at most a single derivative per field. Thus, the boundary term \eqref{fpbdy} has made it possible to define the decoupling limit in terms of the fully diagonalized, physical helicity modes, and has also guaranteed that these modes obey a well-defined variational problem.~This latter property is analogous to what the Gibbons-Hawking term does in general relativity. 

\subsection{Boundary terms for the full theory}
\label{btnl}
\noindent
Fierz-Pauli theory of a free spin-2 particle has a unique consistent extension to non-linear ghost-free massive gravity, discussed in Sec.~\ref{w/obound}.~Furthermore, we have found in the previous sections than consistency of FP theory on a manifold with a boundary itself requires a novel boundary term.~One can then plausibly assume, that also this free field theory boundary term has generalization to the case of nonlinear massive gravity.~In this subsection, we will present such a generalization.

\vskip 0.15cm
\noindent
First we note that our dynamical bulk spacetime (parametrized by the coordinates $X^M$) can be usefully thought of as a spacetime-filling `brane', floating in a \textit{flat} target space parametrized by the `coordinates' $\Phi^I$. The tensor $\mathbb{f}_{MN}$ that enters into the definition \eqref{K} of the building block  $\mathbb K^M_N$ of the graviton's mass/potential terms is then nothing other than the induced metric on that `brane'.~(From this perspective it is intuitively clear that any metric $g_{MN}$, related to $\eta_{MN}$ by a diffeomorphism is generically a solution to the equations of motion of the theory.)
The flat background is characterized by the expectation values \eqref{scalavevs} for the metric and the auxiliary scalars in Cartesian coordinates $X^M$ (obviosuly, this should also be supplemented by $\chi =0$).

\vskip 0.15cm
\noindent
Suppose now that the physical boundary is described by an embedding equation, written in Cartesian coordinates as 
\be
X^5 - \Theta(X^0,\dots,X^3) = 0\,,
\ee
with $\Theta$ some function of the first four of the $X^M$.~In our particular case of interest where the boundary sits at $z=0$, we have $\Theta(X^0,\dots,X^3) = 0$.~Furthermore, the presence of the boundary gives rise to a preferred foliation of the physical vacuum manifold, and we will find it convenient to introduce an analogous foliation in the auxiliary $\Phi$-space.~To this end, we introduce the following \textit{scalar}  of $X^M$-diffeomorphisms
\be
\label{Sigma}
\Sigma(\Phi) \equiv \Phi^5 - \Theta(\Phi^0,\dots, \Phi^3)\,,
\ee
which, according to \eqref{scalavevs}, has the vacuum expectation value $\langle\Sigma \rangle$ equal to zero.~Generically, there is no gauge in which the \textit{fluactuations}  of $\Sigma$ vanish as well, however.~In the case with vanishing $\Theta$ which we are primarily interested in, eq.~\eqref{Sigma} reduces to 
\be
\label{sigmazeq0}
\Sigma =\Phi^5 =  z +\frac{1}{M_5^{3/2} m}\,V^z , 
\ee
and the perturbed boundary value of $\Sigma$ reads (in terms of the canonically normalized fields):
\be
\Sigma | =\frac{1}{M_5^{3/2}}\,\chi + \frac{1}{M_5^{3/2} m}\, V^z\big| \,.
\ee
Again, $\Sigma$ being a spacetime scalar, this is clearly (linearly) gauge invariant -- see the discussion in the previous section.~For further convenience, we will also define the projection $\phi^I$ of the auxiliary scalar multiplet on the $\Phi^5 = 0$ `boundary' of the \textit{internal} space:
\be
\phi^I = \(\delta^I_J - \mathbb{n}^I \mathbb{n}_J\) \Phi^J\,.
\ee
The vector $\mathbb{n}$ here denotes the unit normal to the $\Phi$-space `boundary', with its only non-zero entry being $\mathbb{n}_5 = 1$.~This means that the projected multiplet has the following form: $\phi^a =\delta^a_I\,\Phi^I~(a = 0,\dots,3)$ and $\phi^5 = 0$.~Being spacetime scalars, $\Sigma$ and $\phi^I$ can be used as building blocks of a boundary action.

\vskip 0.15cm
\noindent
One can construct other such building blocks by means of projecting various bulk tensors onto the \textit{physical} spacetime boundary.~One object we will find particularly useful in the following discussion is
\be
\label{phimn}
\phi_{\mu\nu} = \p_\mu X^M \p_\nu X^N \p_M \phi^I \p_N \phi^J\eta_{IJ}\,,
\ee
which is a rank-2 tensor under boundary diffeomorphisms and a scalar under bulk diffeomorphisms.~Notice that it is the projected multiplet $\phi^I$, not $\Phi^I$, that enters into this definition.~The tensor $\phi_{\mu\nu}$ can be used to construct the four-dimensional boundary analog of the building block $\mathbb{K}^M_N$ of the 5D graviton's mass/potential terms:
\be
k^\mu_\nu = \delta^\mu_\nu - (\gamma^{\mu\sigma} \phi_{\sigma\nu})^{1/2}\,,
\ee
where $\gamma_{\mu\nu}$ is the induced metric on the physical boundary, defined in eq.~\eqref{indmet}.
Let us note two defining properties of $k^\mu_\nu\,$.
\begin{itemize}
\item
First, in the gauge \eqref{ug} we are working in and upon decomposing the auxiliary scalars as in eqs.~\eqref{h0} and \eqref{h1}, $k^\mu_\nu$ has the following expansion to the \textit{linear} order in the fields  
\be
\label{prop1}
k^\mu_{\nu}\big|_{\rm lin} = \frac{1}{2 M_5^{3/2}}\bigg[h^\mu_\nu -\frac{1}{m}(\p^\mu V_\nu + \p_\nu V^\mu)\bigg ].
\ee
All indices here have been raised/lowered with the help of the Minkowski metric and we have written this formula in terms of the canonically normalized fields.

\item
Second, $k^\mu_{\nu}$ obeys a \textit{fully non-linear} relation, analogous to eq.~\eqref{propofk} satisfied by its 5D counterpart $\mathbb{K}^M_N$
\be
\label{kmn}
k^\mu_\nu\big |_{h_{MN}=A^M=\chi=0} = \frac{\p^\mu\p_\nu\Pi}{\Lambda_{7/2}^{7/2}}\,.
\ee
Perhaps we should note here that had we used the full scalar multiplet $\Phi^I$ instead of the projected one $\phi^I$ in the definition of $\phi_{\mu\nu}$ (eq.~\eqref{phimn}), this relation would not hold.
\end{itemize}
With these properties of $k^\mu_\nu$ in mind, we'd like to consider the following term in the boundary action, invariant under both the bulk and the boundary (fully non-linear) diffeomorphisms:\footnote{Notice that this term breaks the internal Lorentz group \eqref{internagroup} of the bulk theory.~One could in principle restore this symmetry by promoting the normal $\mathbb{n}^I$ to a \textit{dynamical} vector of the internal group $n^I\to \Lambda^I_{~J}\, n^J$.~This obviously leads to a  different theory with extra degrees of freedom and we won't pursue this possibility in the present work.}
\begin{align}
\label{bt2}
\begin{split}
\int_{\p \it{M}} d^4 x\,\mathcal{L}_2&=\frac{M_5^3 m^2}{6}\int_{\p \it{M}} d^4 x\,\sqrt{-\gamma}\, \Sigma(\Phi) \, \epsilon_{\mu\alpha\rho\sigma}\epsilon^{\nu\alpha\rho\sigma}\, k^\mu_\nu \\ &= \frac{M_5^3 m^2}{6} \int_{\p \it{M}} d^4 x\,\sqrt{-\gamma}\, \Phi^5\,  \epsilon_4\epsilon_4 \times  k\times \mathbb{1}^3\,.
\end{split}
\end{align}
In the second equality we have used eq.~\eqref{sigmazeq0} and (the four-dimensional analog of) the notation defined in \eqref{Vs}.~At the \textit{lowest order} in field fluctuations, the action in \eqref{bt2} reduces to: $2\,\mathcal{L}_2 = -m^2 (\chi + V^z\, m^{-1})(h_4 - 2\, \p_\mu V_\mu\, m^{-1})$, which is precisely our boundary term \eqref{fpbdy} for Fierz Pauli theory, covariantized \textit{a la} \stu and written in terms of the canonically normalized fields (see also the discussion in the previous section).~That the quadratic piece of $\mathcal{L}_2$ coincides with $\mathcal{L}_{\rm FP}^{\rm bdy}$ is a direct consequence of the property \eqref{prop1} of our construction. 
Furthermore, as we have  seen in the previous section, one virtue of adding the given boundary term is that it eliminates the effects of the $(\p_{\mathbb{5}}^2\Pi)^2$ (total bulk derivative) self-interactions of the helicity-0 mode,  which become singular in the decoupling limit.~If not eliminated, these operators would obstruct the possibility of taking the $\Lambda_{7/2}$-decoupling limit as soon as boundaries are introduced to the background spacetime.
\vskip 0.15cm
\noindent
This discussion suggests an obvious criterion for constructing the rest of the boundary terms for the most general non-linear theory of ghost-free massive gravity.~Namely, for each of the bulk mass/potential term $U_n$ in \eqref{Vs}, one needs to find a boundary counterpart $\mathbb{B}_n$ that would precisely cancel the helicity-0 self-interactions of the form \eqref{piaction}, originating from $U_n$.~Guided by the previous observations of this section, it is straightforward to write down such a counterpart:
\begin{align}
\label{bulkboundcorr}
\begin{split}
{\rm bulk:}~U_n = \epsilon_5\epsilon_5\times \mathbb{K}^n\times\mathbb{1}^{5-n}\quad \xleftrightarrow{\hspace*{0.5cm}} \quad
{\rm boundary:}~\mathbb{B}_n = -2\, \Phi^5\, \epsilon_4\epsilon_4\times k^{n-1}\times\mathbb{1}^{5-n}\,,
\end{split}
\end{align}
where the bulk and boundary terms contribute to the fully diffeomorphism-invariant action in the following way
\be
\label{actions}
S^{(n)}_{\rm bulk} =  \int_{\it{M}}d^5X\,\sqrt{-g}\,U_n\,, \qquad S^{(n)}_{\rm bdy} =  \int_{\p\it{M}}d^4 x\,\sqrt{-\gamma}\,\mathbb{B}_n\,.
\ee
In a slighly more expanded form, e.g. the cubic boundary action $S^{(3)}_{\rm bdy}$ is:
\be
S^{(3)}_{\rm bdy} =  \int_{\p\it{M}}d^4 x\,\sqrt{-\gamma}\,\mathbb{B}_3 = - 2 \int_{\p\it{M}}d^4 x\,\sqrt{-\gamma}\,\Phi^5\,\epsilon_{\mu\alpha\rho\gamma}\epsilon^{\nu\beta\rho\gamma}\,k^\mu_\nu\, k^\alpha_\beta\,.
\ee

\vskip 0.15cm
\noindent 
One can check, using eqs.~\eqref{h1}, \eqref{propofk}, \eqref{sigmazeq0} and \eqref{kmn}, that the dangerous helicity-0 self-interactions discussed in Sections~\ref{w/obound} and \ref{secdl} do indeed cancel once a given bulk term is supplemented by the corresponding boundary contribution given in eqs.~\eqref{bulkboundcorr} and \eqref{actions}.\footnote{A relation that helps to establish this is: $$\int\,d^5x\,\varepsilon_5\varepsilon_5\times  (\p_{\mathbb{5}}^2\Pi)^n\times \mathbb{1}^{5-n} = - 2 \int_{z=0} d^4x\, \p_z\Pi \,\varepsilon_4\varepsilon_4\times  (\p_{\mathbb{4}}^2\Pi)^{n-1}\times \mathbb{1}^{5-n}\,.$$}
The theory so defined admits a valid short-distance description in terms of the relativistic 5D helicity modes, strongly coupled at the scale $\Lambda_{7/2}$.~Without the boundary terms $\mathbb{B}_n$ in place, on the other hand, the low-energy theory below the scale $\Lambda_{7/2}$ would suffer from a ghost.

\vskip 0.15cm
\noindent
Again, it is one of the special properties \eqref{kmn} of our construction that guarantees that the boundary terms given in \eqref{bulkboundcorr} cancel the effects associated with the $(\p_{\mathbb{5}}^2\Pi)^n$ operators stemming from the graviton's bulk mass/potential terms $U_n$\,.~For example, had we used the full scalar multiplet $\Phi^I$ instead of the projected one $\phi^I$ in the definition of $\phi_{\mu\nu}$ in \eqref{phimn}, eq.~\eqref{kmn} would not be true and the cancellation would not occur.

\vskip 0.15cm
\noindent
We note, finally, that in the case that the spacetime boundary differs from the constant-$z$ hypersurface, the correct massive gravity boundary terms can be obtained from \eqref{bulkboundcorr} and \eqref{actions} by simply replacing $\Phi^5 \to \Sigma(\Phi)$, with everything else unchanged.

\vskip 0.15cm
\noindent
Let us summarize our findings of this section:~to each of the possible diffeomorphism-invariant mass/potential terms \eqref{Vs} for the graviton in 5D, there corresponds a counterpart on the spacetime boundary, which is invariant under both bulk and boundary diffeomorphisms and needs to be included for well-definiteness of the theory.~Eqs.~\eqref{bulkboundcorr} and \eqref{actions} define this correspondence. 

\section{Conclusions}
\label{conclusions}
\noindent
A necessary  \cite{deRham:2010ik} and sufficient \cite{Mirbabayi:2011aa} requirement for a theory of massive gravity to be ghost-free is that the graviton's helicity-0 polarization $\Pi$ be governed by second-order (classical) equation of motion in a certain short-distance limit, known as the decoupling limit.~For \textit{generic} theories of massive gravity this is not the case as the decoupling-limit Lagrangian of such theories contains higher-derivative operators of the schematic form $(\p^2\Pi)^n$.~Fierz-Pauli theory of a free spin-2 particle and its interacting dRTG extension go around the problem by arranging these unwelcome operators to collect into exact total derivatives so that they drop out of the dynamics in infinite spacetime (given that the fields behave well-enough at infinity).~In any spacetime \textit{with a boundary}, however, the issue comes back:~even if the bulk action is from the ghost-free FP/dRGT class, the bulk total derivatives are not eliminated, but integrate to higher-derivative \textit{boundary} operators.~These boundary terms reintroduce higher-derivative contributions to the $\Pi$-equation of motion.~FP/dRGT theories therefore require a proper extension to remain ghost-free, once the background spacetime is endowed with boundaries.~In this paper, we have constructed such an extension.

\vskip 0.15cm
\noindent
We have found that in order to correctly define the Fierz-Pauli theory of a free massive graviton in a spacetime with a boundary, one ought to include a novel boundary term \eqref{fpbdy} in the action.~We have derived this term in three different ways: (i) by requiring that the equation of motion for the boundary bending mode be consistent with the standard bulk Fierz-Pauli equations (Sec.~\ref{FP}) (ii) by considering the Kaluza-Klein modes of a 6D \textit{massless} graviton in the presence of a boundary, with one spatial direction along the boundary compactified on a circle (Sec.~\ref{FP})  and (iii) by requiring that the higher-(total-)derivative operators of the form $(\p^2\Pi)^n$ coming from the bulk cancel, once they are written as boundary terms (Sec.~\ref{secdl}).~The latter property of the theory allows to define the standard decoupling limit in terms of the fully diagonalized, physical helicity modes.~Moreover, the new boundary operator \eqref{fpbdy} guarantees that the helicity-0 graviton $\Pi$ obeys a well-defined variational problem, entering the Lagrangian with at most one derivative per field.

\vskip 0.15cm
\noindent
We have generalized the boundary term \eqref{fpbdy} for Fierz-Pauli theory to the case of  ghost-free nonlinear theories of massive gravity \cite{deRham:2010ik,deRham:2010kj}.~Requiring that the higher-(total-)derivative bulk operators for the helicity-0 graviton be eliminated from the dynamics led us to conclude that for each of the possible mass/potential terms \eqref{Vs} in the bulk one should include a novel operator in the boundary action.~These non-linear boundary terms are given in eqs.~\eqref{bulkboundcorr} and \eqref{actions}; they are invariant under both the bulk and boundary diffeomorphisms and need to be included for well-definiteness of the theory.~Basing ourselves on physical arguments, we have not provided a strict mathematical proof (e.g.~a full-fledged Hamiltonian analysis) that would establish our generalization of the Fierz-Pauli boundary term \eqref{fpbdy} to the case of fully non-linear and diff-invariant massive gravity as the unique one. However, the discussion of Sec.~\ref{btnl} strongly suggests that this generalization is indeed unique.

\vskip 0.15cm
\noindent
There are a number of directions to extend our work in.~First, in the upcoming work \cite{upcoming} we study the consequences of the above-derived boundary terms for the dynamics of massive gravity on AdS$_5$ with a flat boundary -- a setup that leads to an effective (boundary) theory of 4D massive gravity with high cutoff \cite{Gabadadze:2017jom}.~Furthermore, using the machinery of refs.~\cite{Shiromizu:1999wj} and \cite{Binetruy:1999hy}, it would be interesting to study cosmology of the resulting theories, in particular whether or not the setup entails (self-accelerated) solutions with a \textit{de Sitter} boundary.

\subsection*{Acknowledgements}
\noindent
We thank Cedric Deffayet, Claudia de Rham and Andrew Tolley for valuable discussions.~The work of GG was supported in part by NSF grant PHY-1620039.~DP is supported by European Union's Horizon 2020 Research Council grant 724659 MassiveCosmo ERC-2016-COG.~DP would like to thank the Center for Cosmology and Particle Physics at New York University for hospitality during his visit, supported by the Simons Foundation under the program `Origins of the Universe'.

\renewcommand{\em}{}
\bibliographystyle{utphys}
\addcontentsline{toc}{section}{References}
\bibliography{bibliography}

\end{document}